\documentclass[preprint]{aastex}
\usepackage{graphicx}
\begin{document}

\title{Termination of planetary accretion due to gap formation.}

\author{R. R. Rafikov}
\affil{Princeton University Observatory, Princeton, NJ 08544}
\email{rrr@astro.princeton.edu}

\begin{abstract}
The process of gap formation by a growing planetary embryo embedded in a
planetesimal disk
is considered. It is shown that there exists a single parameter 
characterizing this process, which represents the competition between 
the gravitational influence of the embryo and planetesimal-planetesimal 
scattering. 
For realistic assumptions about the properties of the
planetesimal disk and the planetary embryo, a 
gap is opened long before the embryo
 can accrete all the bodies within its region of influence. 
The implication of this result is that the 
embryo stops growing and, thus, large 
bodies formed during the coagulation stage should be less massive 
than is usually assumed. 
For conditions expected at $1$ AU in the solar protoplanetary disk, 
gap formation is expected to occur around bodies of mass $\la 10^{24}$ g.
The effect of protoplanetary radial
migration is also discussed.
\end{abstract}


\section{Introduction.}  \label{intro}

The formation of planets is one of the most complex problems in astrophysics,
involving accumulation of bodies over some $45$ orders of magnitude in mass 
--- from dust grains to giant planets.

One issue which has received 
a lot of attention  is the formation of planetary embryos by 
the accretion of  planetesimals.
From the perspective of dynamics we call an object an embryo 
when it becomes so massive that one can no longer
 describe its behavior by means of simple kinetic theory (in this paper
we use names embryo, protoplanet
 and massive body interchangeably). In other words,
in the presence of embryos the multiparticle distribution function 
cannot be taken as a product of one-particle distribution functions;
the gravitational influence of the embryo is strong enough to affect 
the distribution of planetesimals with which it interacts. 
For example, a gap could form around the embryo.
 To study properties of systems containing 
embryos one must either resort to N-body simulations or try 
to account properly for their influence
 on the underlying planetesimal population and on each other.

In the standard scenario, protoplanets
 grow in orderly (Safronov, 1972)  or runaway fashion
(Wetherill \& Stewart, 1989; Wetherill \& Stewart, 1993) 
by accreting planetesimals from the protoplanetary nebula.  
After the largest bodies become embryos, they open gaps and accretion
slows or stops.
Thereafter these massive bodies evolve more
slowly as gravitational encounters perturb them into crossing orbits and 
violent impacts occur, thus gradually forming more massive objects. 
It is a common
belief now that
 the Earth-type planets and rocky cores of the giant planets were formed
by this two-stage process.

The question which is not very often addressed is where the boundary between 
these two stages occurs. This is an 
 important issue, because the answer tells us  the final
mass of the objects which further evolve through chaotic collisional 
evolution, as well as the number of such bodies, and the conditions 
for which statistical treatments of collisional evolution are valid.

The standard paradigm for determining the embryo mass
(Lissauer, 1987; Weidenschilling et al. 1997)
presumes that 
planetary growth stops when the embryo ``eats up'' all the planetesimals 
within a ``feeding zone'' -- an annulus of radial width of one Hill radius. 
Here the
Hill radius is defined as
\begin{equation}
R_H=a\left(\frac{M_p}{M_c}\right)^{1/3},\label{r_h}
\end{equation}
with $a$ being the distance between the massive body with mass $M_p$ and the
central star with mass $M_c$. If the surface mass density of planetesimals
is $\Sigma_0$, then this ``isolation mass''
is $M_p=M_{is}\sim 2\pi a R_H \Sigma_0$, which means that 
\begin{equation}
M_{is}\sim \frac{(2\pi a^2\Sigma_0)^{3/2}}{M_c^{1/2}}=
1\times 10^{26}~\mbox{g}~
\left(\frac{\Sigma_0 a^2}{2\times 10^{-6}~M_\odot}\right)^{3/2}
\left(\frac{M_c}{M_\odot}\right)^{-1/2}.\label{isol} 
\end{equation}
The estimate of $\Sigma_0 a^2$ is made for $1$ AU 
and is based on standard assumptions for the 
protosolar nebula: $\Sigma_0\approx 20 (a/\mbox{1 AU})^{-3/2}$ g cm$^{-2}$,
implying that $\approx 1$ per cent of the
minimum mass Solar nebula is contained 
in solids (Hayashi 1981). 

The usual assumption is that before
the mass of the largest protoplanet reaches $M_{is}$, the distribution of
planetesimals is basically homogeneous. 
In some cases this is not a reasonable 
assumption. In particular Ida \& Makino (1993)
demonstrated using N-body simulations that a protoplanet could scatter
planetesimals strongly if it is massive enough and, thus, clear a zone 
around it which is free from any solid bodies. The mass required to clear a gap in this way is not simply related
to $M_{is}$.
Kokubo \& Ida (1998) later
emphasized the importance of rapid heating of the planetesimal population 
by the forming planet in slowing down the subsequent accretion of
planetesimals.

The process of clearing a gap in the planetesimal disk 
around a massive body is 
analogous to  gap formation in gaseous disks (Takeuchi et al. 1996),
which results from a competition between viscous spreading of the disk
and gravitational interactions with the protoplanet.
In a planetesimal disk the role of viscosity 
 is played by mutual scattering of planetesimals. One can
easily  estimate the planetary embryo mass determined by 
this process. Let us assume that the planetesimal random velocities are small 
-- the disk is cold. Let $\Omega=(GM_c/a^3)^{1/2}$ be the orbital 
angular velocity.
If $m_0$ is the mass of each of two planetesimal and 
$r_H=a(2m_0/M_c)^{1/3}$ is the corresponding Hill radius,
 then the typical displacement in a close encounter of these planetesimals
on circular orbits separated by $< r_H$
is $\sim r_H$, and the typical random
velocity kick is $\Omega r_H$.
Similarly, planetesimals within $R_H$ from the massive object 
get kicked by $\sim R_H$
with frequency $\sim \Omega (R_H/a)$ (the inverse of the synodic period). 
If they are not able to diffuse back 
this distance during the time interval between kicks by the embryo, then a
 gap forms. 

The ``viscous'' spreading distance is $\sim r_H K^{1/2}$ during one 
synodic period, where $K$ is 
the number of collisions of a given small body with other planetesimals between
consecutive approaches to the massive body. Assuming that the thickness
of the planetesimal disk is $\sim r_H$, one can easily see that 
 $K \sim r_H^2(\Sigma_0/m_0)(a/R_H)=
a^2 \Sigma_0 /(m_0 M_p M_c)^{1/3}$.
 This implies that a gap in the planetesimal 
disk opens when $R_H^2\ga r_H^2 K$, or when  
\begin{equation}
\frac{M_p M_c^{1/3}}{f(v/\Omega r_H)
\Sigma_0 a^2 m_0^{1/3}}\ga 1,\label{lam0}
\end{equation}
or, alternatively, when 
\begin{equation}
M_p>M_{crit}\sim f(v/\Omega r_H)\Sigma_0 a^2
\left(\frac{m_0}{M_c}\right)^{1/3}.
\label{critestim}
\end{equation}
Here $f(v/\Omega r_H)$ is a dimensionless 
function characterizing the effect
of the planetesimal velocity dispersion $v$; we
expect $f(x)\sim 1$ for $x\la 1$ and $f(x)\sim x^{-2}\ln x$ for 
$x\gg 1$ (see \S \ref{randmot}). 
If the mass  given by equation (\ref{critestim})
is smaller than the isolation mass given by equation 
(\ref{isol}), the accretion stops because the protoplanet forms a gap, rather 
than because it consumes all the bodies in its feeding zone.
Assuming typical values for protoplanetary disks one can get
\begin{equation}
M_{crit}\approx 4\times 10^{23}~\mbox{g}~f(v/\Omega r_H)~
\left(\frac{\Sigma_0 a^2}{2\times 10^{-6}~M_\odot}\right)
\left(\frac{m_0}{10^{21}\mbox{g}}\right)^{1/3}
\left(\frac{M_c}{M_\odot}\right)^{-1/3}.
\label{critest}
\end{equation}
It is obvious from this estimate that gap formation could be very important
in slowing down planetary accretion.

In this paper we analytically 
study the process of clearing a gap around a massive body in a planetesimal 
disk. 
We use an approach to treating the surface density evolution
that was first developed by Petit \& H\'enon
in their seminal series of papers (1987a, 1987b, hereafter PH, 1988). 
In \S \ref{equations} and Appendix \ref{ap1} we derive a generalized 
form of their evolution equation, including the 
fluxes produced by the protoplanet 
and those generated by
mutual gravitational perturbations between the planetesimals of 
the swarm.

In \S \ref{res} we describe the  solutions of the evolution 
equations for cold planetesimal disks,
and compare our results with those obtained using 
N-body simulations. We comment on the applicability of our findings 
to the planet formation in the early Solar System, 
and describe briefly the relation
between the surface density and planetesimal velocity dispersion
evolution in \S \ref{disc}.

\section{Derivation of the general equation.}\label{equations}

All the following calculations assume a Keplerian disk, 
 although they could be easily extended to the case of
an arbitrary rotation law.

We consider a disk of bodies (we will refer to them as planetesimals, 
but these could
be other bodies, such as planetary ring particles)
 with $N(m,r,t)dm=\Sigma(m,r,t)dm/m$ being the surface
number density of particles with mass between $m$ and $m+dm$, 
whose guiding centers move
at a distance  $r$ from the
central body. It is important to keep in mind that 
$r$ is the guiding center radius rather than the instantaneous radius.
The  instantaneous surface number density can only be obtained if
$\Sigma(m,r,t)$ is supplemented by the random velocity distribution of 
planetesimals.

We also assume that a single
massive body with mass $M_p$ moves on a circular orbit
 in this planetesimal swarm
(we take orbit to have a fixed radius 
and we comment later on the effects of migration)
and we assume that its mass is much larger than the masses of the individual
swarm particles.
The mass of the central body is  $M_c$ and the distance
of the planet from the central body is $a$. 
We will also use the relative
 masses of the
bodies with respect to $M_c$: $\mu_p=M_p/M_c$ for the 
planet and $\mu=m/M_c$
for the planetesimals with mass $m$. 

The interactions between particles in the gravitational field of a central body
are described by Hill's equations, which are valid in the limit 
$\mu, \mu_p \ll 1$,
which is  always true in problems which we will study.
It was demonstrated by H\'enon \& Petit (1986) that
in this case the motion of nearby gravitationally interacting particles 
can be separated
into center-of-mass motion, which is invariant during the interaction, and
relative motion. 
If one defines new dimensionless  
coordinates  where all the distances and relative velocities are normalized
by $a(\mu_1+\mu_2)^{1/3}$,
then the equations of relative motion of particles 1 and 2 do not depend on
their masses in these coordinates.
Let us set  $h$ to be the distance between the guiding centers of interacting
particles in these coordinates and
 $P(h,\Delta h)d\Delta h$
to be the probability of having a change in $h$ in the range $(\Delta h,
\Delta h +d\Delta h$
in an encounter. 
Then $P(h,\Delta h)$
 does not depend upon the masses of particles involved in a collision,
but does depend on the 
random velocity distribution function 
of the planetesimals.

In Appendix \ref{ap1} we derive the general equation of the 
surface density evolution,
which in many aspects parallels the derivation of equation (44) in PH.
Let us set $N_i=N(m_i,r,t)$. Let $A=(r/2)(d\Omega/dr)$ be the
 function determining
the local shear, $A=-(3/4)\Omega$ for a Keplerian rotation law. Then the 
surface density
evolution is given by
\begin{eqnarray}
\frac{\partial N_1(r)}{\partial t}=-2|A|a^2\int\limits_0^{\infty}dm_2
(\mu_1+\mu_2)^{2/3}
\int\limits_{-\infty}^{\infty}dh |h|
\Bigg\{N_1(r)N_2[r-(\mu_1+\mu_2)^{1/3} r h]\nonumber \\-\left.
\int\limits_{-\infty}^{\infty}d(\Delta h)P(h,\Delta h)
N_1[r+D(\Delta h)]N_2[r+D(\Delta h)-(\mu_1+\mu_2)^{1/3} a h]
\right\},\label{geneq}
\end{eqnarray}
 where
\begin{equation}
D(\Delta h)=-\frac{\mu_2 a}{(\mu_1+\mu_2)^{2/3}}\Delta h.\label{D}
\end{equation}
Note that the factor $r$  is 
replaced with $a$ in equations (\ref{geneq}),(\ref{D}), 
where it is appropriate,
 because at this level of approximation there is 
no difference between them, since they are both much larger than the 
Hill radius. 

This  differs from the equation derived in PH because
it does not  assume that surface density varies slowly
 on scales of the order of the 
Hill radius.
If we made this assumption and expanded $N(m,r,t)$ up to the second 
order in $h$ locally
we would reduce equation (\ref{geneq})  to the one
derived in PH, equation (44), which we reproduce here:
\begin{eqnarray}
\frac{\partial N_1}{\partial t}= |A|r^4
\int\limits_0^{\infty}dm_2\left[2I_1\mu_2(\mu_1+\mu_2)^{1/3}
\frac{\partial}{\partial r}\left(N_1
\frac{\partial N_2}{\partial r}\right)+
I_2\frac{\mu_2^2}{(\mu_1+\mu_2)^{2/3}}
\frac{\partial^2(N_1 N_2)}{\partial r^2}
\right].
\label{ph}
\end{eqnarray}
Here $I_1$ and $I_2$ are dimensionless
 moments of the probability distribution
$P(h,\Delta h)$:
\begin{eqnarray}
I_1\equiv\int\limits^{\infty}_{-\infty}|h|hdh 
\int\limits^{\infty}_{-\infty} d(\Delta h)
\Delta h P(h,\Delta h)=\int\limits^{\infty}_{-\infty}|h|hdh \langle 
\Delta h\rangle,\label{int-s1}\\
I_2\equiv\int\limits^{\infty}_{-\infty}|h|dh 
\int\limits^{\infty}_{-\infty} d(\Delta h)
(\Delta h)^2 P(h,\Delta h)=
\int\limits^{\infty}_{-\infty}|h|dh \langle 
(\Delta h)^2\rangle,\label{int-s2}
\end{eqnarray}
and symmetry dictates that
\begin{eqnarray}
\int\limits^{\infty}_{-\infty}|h|dh 
\int\limits^{\infty}_{-\infty} d(\Delta h)
\Delta h P(h,\Delta h)=0.
\end{eqnarray}

For a cold disk  it was demonstrated by PH that
\begin{equation}
I_1=-3.07,~~~I_2=17.72.\label{Ivalues}
\end{equation}

We can now easily include
 the effect of a massive body on the surface density
evolution.
To do this we take surface density
 to consist of two parts: one representing a continuous
distribution of small masses, corresponding to planetesimals, and another
arising from the massive body. One
 can write the contribution from the embryo in the following form:
\begin{equation}
N_{em}(m,r,t)=\frac{1}{2\pi a}\delta(m-M_p)\delta(r-a).\label{formdistr}
\end{equation}
We neglect migration of the embryo, so we assume $a$ is fixed.

Substituting (\ref{formdistr}) into (\ref{geneq}) and assuming
that the embryo is much more massive than any
of the planetesimals, $M_p\gg m$, we get
\begin{eqnarray}
\frac{\partial N_1}{\partial t}=-2 A a^2\int\limits_0^{\infty}dm_2
(\mu_1+\mu_2)^{2/3}
\int\limits_{-\infty}^{\infty}dh |h|\Bigg\{
N_1(r)N_2[r-(\mu_1+\mu_2)^{1/3}ah]\nonumber \\-\left.
\int\limits_{-\infty}^{\infty}d(\Delta h)P(h,\Delta h)
N_1[r+D(\Delta h)]
N_2[r+D(\Delta h)-(\mu_1+\mu_2)^{1/3} a h]\right\}\\
-\frac{A}{\pi a}\left\{N_1(r)|r-a|-\frac{1}{\mu_p^{1/3} a}
\int\limits^{\infty}_{-\infty}
dr_1 N_1(r_1)|r_1-a|P\left(\frac{r_1-a}{a\mu_p^{1/3}},
\frac{r-r_1}{a\mu_p^{1/3}}\right)\right\}.\nonumber
\label{gen}
\end{eqnarray}

We can make other simplifications
taking the following into account.
In our particular problem the relevant length scale for any structure is
the Hill radius of the massive body $R_H$. Planetesimals 
can get kicks when interacting with the massive body which change 
their guiding centers  by $\sim R_H$. At the same time
mutual interactions between the planetesimals are unable to produce such large
displacements (since $\mu_p\gg\mu_i$). One can thus assume that for
planetesimal-planetesimal interactions the surface 
density varies only slowly on the scale $r_H\ll R_H$,
and
locally expand the first part of the r.h.s. of equation (\ref{geneq})
in a Taylor series in $h$, as was
done in equation (\ref{ph}). At the
same time  the second part of the r.h.s., representing the interaction
with the large body, cannot be simplified in a similar way.

We also make some additional
changes: we move the origin of $r$ to $a$ (simply set $r-a=r^\prime$) and
switch from $r^\prime$ to a dimensionless distance from the
planetary embryo $H=r^\prime/(\mu_p^{1/3}a)$. Then we get 
\begin{eqnarray}
\frac{\mu_p^{2/3}}{A a^2}\frac{\partial N_1(H)}{\partial t}=
\int\limits_0^{\infty}dm_2\left[2I_1\mu_2(\mu_1+\mu_2)^{1/3}
\frac{\partial}{\partial H}\left(N_1
\frac{\partial N_2}{\partial H}\right)+
I_2\frac{\mu_2^2}{(\mu_1+\mu_2)^{2/3}}
\frac{\partial^2(N_1 N_2)}{\partial H^2}
\right]\nonumber\\
-\frac{\mu_p}{\pi a^2}\left[N_1(H)|H|-
\int\limits^{\infty}_{-\infty}dH_1~ N_1(H_1)|H_1|
P(H_1,H-H_1)\right].
\label{lin}
\end{eqnarray}
 
In deriving this form of the evolution equation 
we only assumed that $\mu_p\gg\mu_1$. So, this nonlinear
integro-differential equation can adequately describe the evolution  of
the surface density of planetesimals in the disk-protoplanet system.

\subsection{Single mass planetesimals.}

The constituent bodies of 
planetesimal disks are likely to have quite a broad range of masses. 
However, right now we are going to concentrate on
the simple case of single mass planetesimals, that is we assume
all planetesimals to have a unique mass $m_0$.
Then  $N(m,H)=\sigma(H)(\Sigma_0/m_0)\delta(m-m_0)$, where $\Sigma_0$ is the
surface  mass density of particles at infinity which we take to be a
reference value
(it follows then that $\sigma(\infty)=1$). Substituting this assumption into
equation (\ref{lin}) and performing an integral over $m_2$ one
obtains that
\begin{eqnarray}
\frac{1}{I}
\frac{\partial \sigma}{\partial \tau}=
\frac{\partial^2 \sigma^2}{\partial H^2}
-\lambda\left[\sigma(H)|H|-
\int\limits^{\infty}_{-\infty}dH_1~\sigma(H_1)|H_1|
P(H_1,H-H_1)\right],
\label{1pop}
\end{eqnarray}
where $\mu_0=m_0/M_c$ and 
\begin{equation}
I\equiv I_1+\frac{I_2}{2}=\frac{1}{2}\int\limits_{-\infty}^{\infty}|h|\langle
2h\Delta h +(\Delta h)^2\rangle dh.\label{Igen}
\end{equation}
The new time variable $\tau$
is defined as
\begin{equation}
\tau=\frac{t}{t_0},~~~~\mbox{where}~~~~ t_0=
\frac{\mu_p^{2/3}M_c}{2^{1/3}A \mu_0^{1/3}\Sigma_0 a^2},\label{taudef}
\end{equation}
 and [cf. eq. (\ref{lam0})]
\begin{equation}
\lambda=\frac{M_p}{2^{1/3}\pi \Sigma_0 a^2 I \mu_0^{1/3}}.\label{crit}
\end{equation}
One should notice that the first term in the r.h.s. of equation (\ref{1pop})
 and the expression in brackets are both dimensionless; all of the 
dimensional information is hidden in $\lambda$ and $\tau$.

For a cold disk (rms velocity dispersion 
of planetesimals in $r$-direction $v_r\ll\Omega r_H$) 
we obtain using (\ref{Ivalues}) that 
\begin{equation}
I=5.79,\label{Icold}
\end{equation}
and, thus,
\begin{equation}
\lambda=0.0436~\frac{M_p}{\Sigma_0 a^2 \mu_0^{1/3}},~~~~~~
v_r \ll\Omega r_H.\label{coldcrit}
\end{equation}

For a hot disk 
($v_r\gg\Omega r_H$) one has 
[see the discussion after equation (\ref{I}) in \S \ref{randmot}]
\begin{equation}
I=29.8~\frac{\Omega^2 r_H^2}{v_r^2}\ln\Lambda,~~~
\mbox{with}~~~
\Lambda\sim\left(\frac{v_r^2}{\Omega^2 r_H^2}\right)^{3/2}.
\label{Ihot}
\end{equation}
 It is assumed here that 
the ratio of vertical to radial random velocity dispersions in 
a planetesimal disk is equal to $0.5$. Using (\ref{Ihot}) one finds 
that
\begin{equation}
\lambda=0.0085~\frac{M_p}{\Sigma_0 a^2 \mu_0^{1/3}}
\frac{v_r^2 }{\Omega^2 r_H^2}
\left(\ln\Lambda\right)^{-1},
~~~~~~v_r\gg\Omega r_H.\label{hotcrit}
\end{equation}
In the intermediate regime, when $v_r\sim\Omega r_H$,
there is no analytic expression for $I$ and one has to interpolate
between the two asymptotic behaviors given by (\ref{Icold}) and (\ref{Ihot}). 

The parameter $\lambda$ 
quantifies the influence of the planetary perturbations on the
uniformity of the planetesimal disk.
The first term on the r.h.s. of (\ref{1pop}) describes the nonlinear 
diffusion of particles due to mutual gravitational scattering,
and tends to iron out any initial inhomogeneities. The second term 
 represents the effect of the planet, which tends to carve out a gap
 in the distribution of  planetesimals. The steady state
originates when these two effects balance each other.
 
When the protoplanetary mass is
large, $\lambda$
is also large and the expression in brackets dominates the evolution.
It leads to gap formation.
On the contrary, if the planetary mass is small we can neglect the
corresponding term  in the r.h.s. of (\ref{1pop})
and obtain a nonlinear diffusion equation, which drives the planetesimal
distribution towards a homogeneous state.
So, one
can say that a 
gap (or at least a significant depression in the surface density 
of the planetesimals) is formed when $\lambda\ga 1$.
From equations (\ref{1pop}) - (\ref{crit}) one can derive the 
characteristic time 
required for a gap to form when $\lambda\ga 1$:
\begin{equation}
t_{open}=\frac{t_0}
{I\lambda}=\frac{\pi}{|A|\mu_p^{1/3}}=\frac{2T}{3\mu_p^{1/3}}
\approx 400~T~\left(\frac{M_p}{10^{25}~\mbox{g}}\right)^{-1/3}
\left(\frac{M_c}{M_\odot}\right)^{1/3},\label{topen}
\end{equation}
where $T=2\pi/\Omega$ is the orbital period of the embryo. Note that 
$t_{open}$ is 
approximately the synodic period of a 
body at $R_H$ from the embryo and, thus,  is 
independent of the planetesimal mass $m_0$, the surface density
$\Sigma$, and the numerical factor $I$.

In \S \ref{res} we confirm these arguments by solving equation
(\ref{1pop}) numerically.

\section{Numerical results.}\label{res}

\subsection{Solution of the equation of evolution.}

\begin{figure}
\vspace{20.cm}
\includegraphics{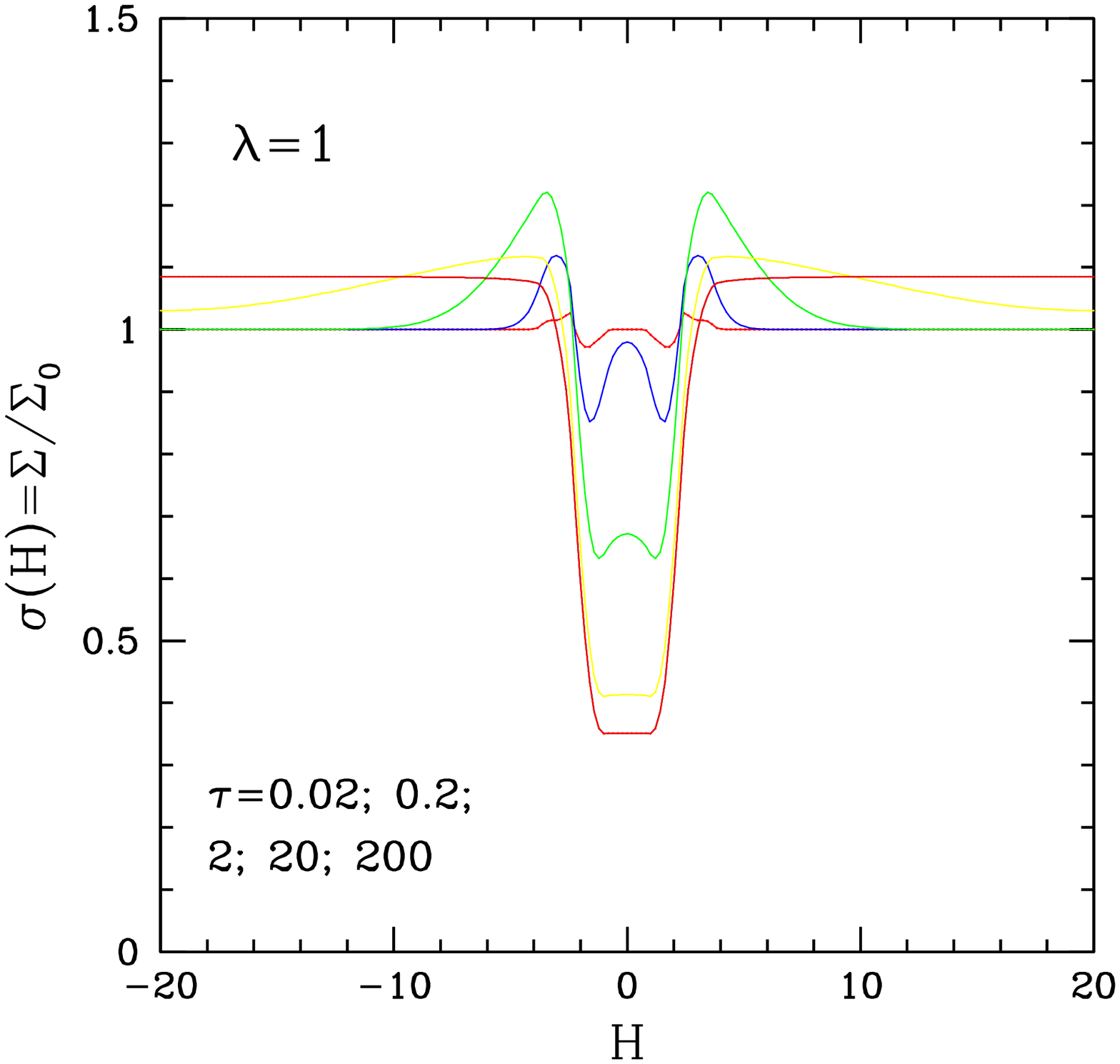}
\includegraphics{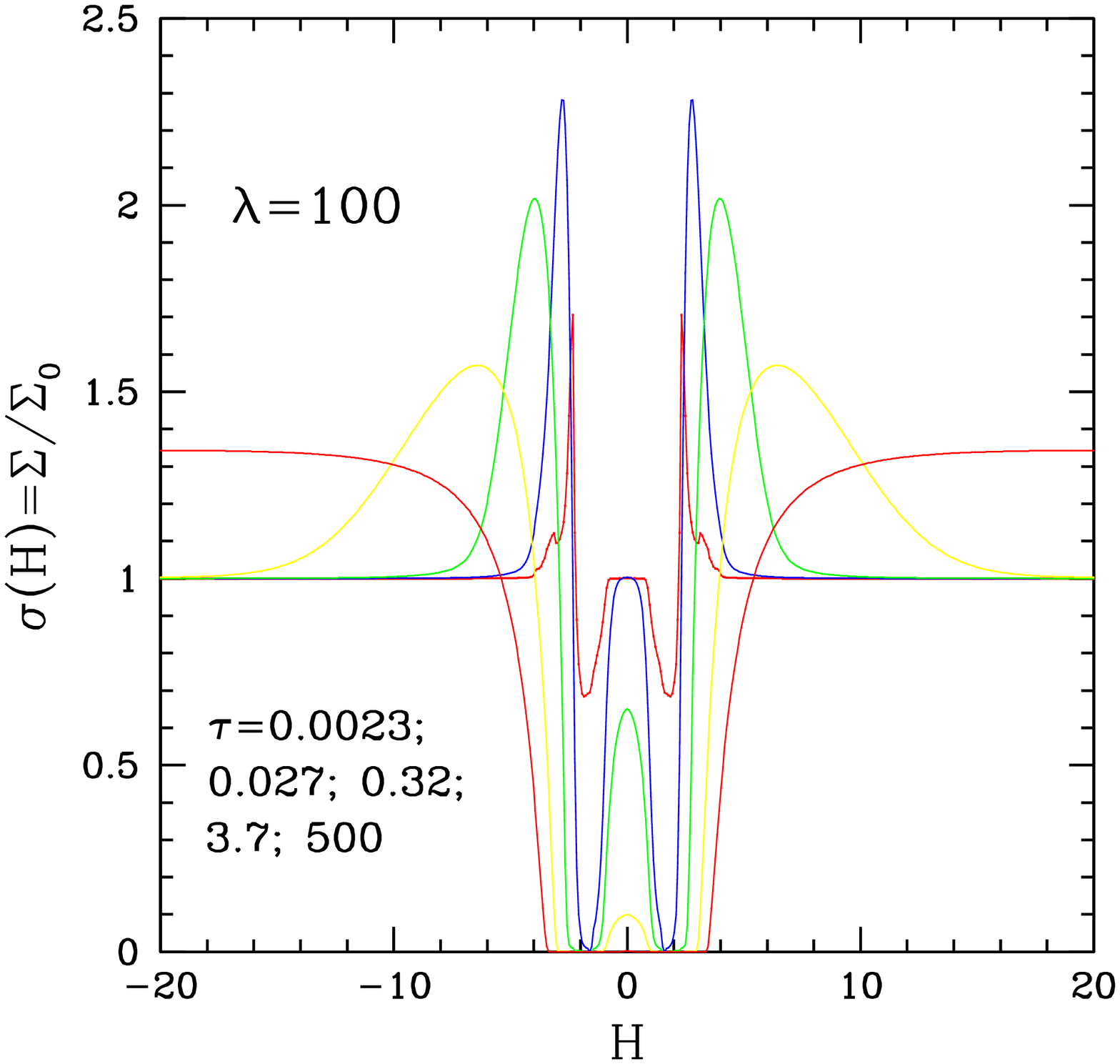}
\caption{
The time evolution of the surface density in a cold planetesimal disk
with a single massive body at
$H=0$, for two values of the parameter
$\lambda$ defined in equation (\ref{crit}). 
In the {\it top panel} the case $\lambda=1$ is described,
in the {\it bottom panel} we consider $\lambda=100$. The dimensionless
time $\tau$ is indicated on the panels; larger $\tau$ corresponds to
a deeper gap. Notice the presence of a bump at the center of 
 a forming gap which is due to  particles in horseshoe orbits
near the massive body. The increase in asymptotic surface density at late 
$\tau$ is an artifact of the use of periodic boundary conditions
at $H=\pm 20$, which forces $\int^{20}_{-20}\sigma(H)dH$ to be conserved.}
\label{fig:evolution}
\end{figure}

We solved equation (\ref{1pop}) in a simplified setting, in which we 
completely neglect the velocity evolution of the planetesimal population.
Thus, we assume the integral $I$ embodying all the kinetic properties of 
planetesimals to be fixed in time. For simplicity we set $I=1$;
this only affects the timescale of the gap formation by a constant factor 
and does not change the evolution at all. 

We also assume that planetesimals have very small random motion
on the scale of the embryo's Hill radius:
$v\ll \Omega R_H$. This simplifies our
treatment a lot, because in this case scattering is deterministic so that 
$P(h,\Delta h)=\delta(\Delta h -
h^\prime(h)+h)$.  The behavior of the function $h^\prime(h)$ 
which gives the final semimajor axis difference as a function of the initial
difference was
described in detail by PH. For the sake of convenience we reproduce this 
dependence in Appendix \ref{ap2}. 
Also, in this approximation, the instantaneous surface number 
density is given simply by $\sigma(H)$, because the guiding center 
and instantaneous radii coincide.
The assumption of a cold disk might be 
reasonable in  cases such as the early stages of  
gap clearing in a planetesimal disk, when scattering
by the embryo could be in the 
shear-dominated regime, or all the time in dense planetary 
rings (Petit \& H\'enon 1987a).

We solved equation (\ref{1pop}) with  periodic boundary conditions, 
simply assuming
that $\partial \sigma/\partial H=0$ at $H=\pm L$. The functional form
of $P(h,\Delta h)$ is given by equation (\ref{phscat}).
We usually assume $L=20.0$ here (in units of the Hill radius of planet) and
take the initial surface number density to be constant: $\sigma(H,0)=1$.

In Figure \ref{fig:evolution} we show the evolution of the surface density
with time [we use the dimensionless time $\tau$ given by equation 
(\ref{taudef})] for  
$\lambda=1$ and $100$. In the first case a gap is never actually formed and 
in the steady state there is only a density depression around the planet.
Thus the particles can be still accreted by the protoplanet, but the efficiency
of this process is reduced.

In the case $\lambda=100$ the gap is formed very quickly, which is in 
general  agreement with the expected
timescale for gap formation ($\tau_{open}\sim 
\lambda^{-1}$ in this case), 
although it takes some time after that for the 
density distribution to settle to a steady state. 

In both cases one should notice 
a bump inside the gap which decays with time. It corresponds to the 
horseshoe orbits in the immediate vicinity of an embryo.
This is in agreement with  Monte-Carlo and N-body simulations
performed earlier (Petit \& H\'enon, 1988; Tanaka \& Ida, 1997; Spahn \&
Srem\v{c}evi\'c 2000).

\begin{figure}
\vspace{9.cm}
\includegraphics{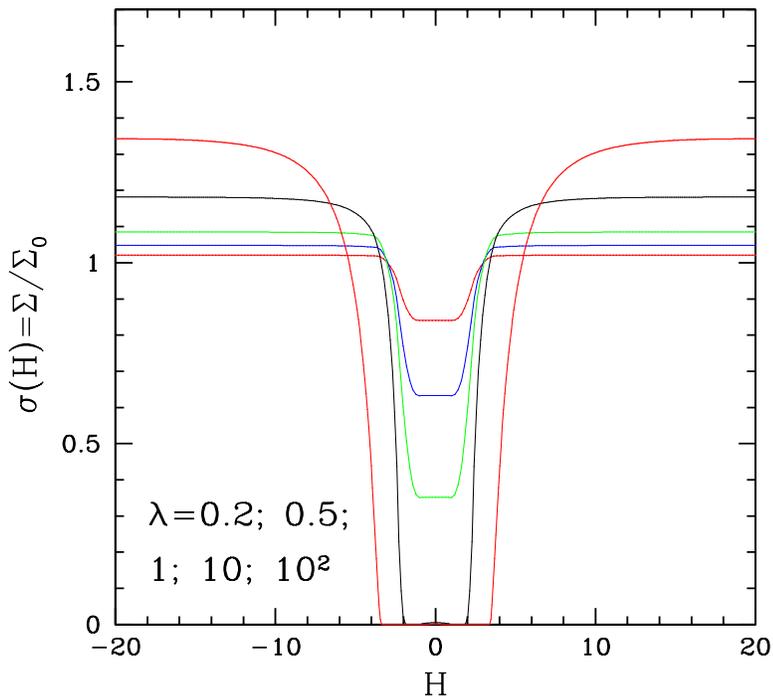}
\caption{
The final distribution of the surface density for a cold disk
with a single massive body, for several values of the parameter
$\lambda$: $0.2; 0.5; 1; 10; 100.$ The higher values of 
$\lambda$ correspond to progressively deeper gaps in the Figure. The increase
in asymptotic surface density at late $\tau$ is an artifact of the use
of periodic boundary conditions. }
\label{fig:steadystate}
\end{figure}

In Figure \ref{fig:steadystate} we show the final state of the surface 
density for several values of $\lambda$, so that one can see that gap gets
deeper and wider as $\lambda$ increases. It also looks like the condition 
$\lambda=1$ is a reasonable approximate criterion for gap formation.

\subsection{Comparison with N-body simulations.}

We may compare our analytical results 
from equation (\ref{crit}) with
N-body simulations by Ida \& Makino
(1993, hereafter IM93) and  Tanaka \& Ida (1997, hereafter TI97).

Figure 3 of IM93 and Figure 1 of TI97 show well developed gaps
in a gas-free planetesimal disk. The gaps seen in 
N-body simulations are never clean because random motion of planetesimals 
is naturally included, and this permits some of them to be present in the gap.
The parameters used in the production of these
Figures correspond to $\lambda\approx 25/I$ in  the first case
while in the second
$\lambda\approx 60/I$. In both cases the velocity dispersion of planetesimals 
is large ($ v /\Omega r_H \sim 20-70$) 
and we expect $I\ll 1$ (see \S \ref{randmot}), 
meaning that $\lambda\gg 1$ and the
condition for gap formation in the distribution of guiding centers of 
planetesimals should be fulfilled. 
 It was also demonstrated by TI that 
gas drag could clear the gap of residual high-velocity planetesimals 
and thus stop accretion completely.

In Figure 6 of IM93 there is shown 
a sequence of scenarios  for
different ratios of the $M_p/m_0$ --- planet to planetesimal masses.
In the case $M_p/m_0=10$, when the gap is barely seen at all,
$\lambda\approx 2.2/I$; since planetesimals are not strongly heated,
presumably $\lambda\la 1$. For $M_p/m_0=30$, when the
gap becomes pronounced,
$\lambda\approx 6.6/I$ and $\lambda\sim 1$. Finally for $M_p/m_0=100$,
when the gap is quite significant, $\lambda\approx 22/I$; heating
also becomes important and $\lambda\gg 1$.
In addition, these results confirm our prediction below  in
\S \ref{randmot} that for $\lambda \sim 1$, 
when gap starts to form,
planetary perturbations begin to dominate planetesimal random velocity 
stirring within
$\sim R_H$ of the planet.

This comparison shows that estimates of gap formation 
based on the  parameter $\lambda$
are qualitatively consistent with N-body simulations.

\section{Discussion}\label{disc}

\subsection{Applications.}

We have established the usefulness of the parameter $\lambda$ in describing
gap formation. Now we are going to use it to determine the mass
of the body which could open a gap in a planetesimal disk. This is simply
done by setting $\lambda=1$. 
We rewrite this condition as 
\begin{equation}
M_{crit}=2^{1/3}\pi I \Sigma_0 a^2 
\left(\frac{m_0}{M_c}\right)^{1/3},\label{critmass}
\end{equation}
where $M_{crit}$ is the planet mass for which $\lambda=1$.
One can easily see that this value of critical planetary mass coincides
with our simple estimate in equation (\ref{critestim}) if we set
$f(v/\Omega r_H)=2^{1/3}\pi I$.

As we mentioned in \S \ref{intro}, it is usually assumed that 
accretion stops when planet hoovers up
a zone in the planetesimal disk with width equal to its Hill
radius [eq. (\ref{isol})]. 
Then the ratio of the two critical masses is
\begin{equation}
\frac{M_{crit}}{M_{is}}=\frac{I}{2^{7/6}\pi^{1/2}}
\left(\frac{m_0}{\Sigma_0 a^2}\right)^{1/3}
\left(\frac{M_c}{\Sigma_0 a^2}\right)^{1/6}.\label{massratio}
\end{equation}

We expect the total mass of the circumstellar disk
to be  of order $0.1-0.01$ of the mass of the central star $M_c$
(Osterloh \& Beckwith 1995; Mannings \& Sargent 2000).
For the protosolar nebula it is often assumed that surface density of gas at
$1$ AU is $\sim 2000$ g cm$^{-2}$ (Hayashi 1981). 
The mass fraction of heavy elements, which contribute to 
solid body formation in this disk, is $\sim 0.01$.  
Using this information we can estimate that $M_c/(\Sigma_0 a^2)
\sim 10^{5}-10^{6}$ at $1$ AU and the corresponding factor
does not contribute a lot to the ratio in (\ref{massratio}) 
(it varies roughly from $7$ to $10$).
The factor $2^{-7/6}\pi^{-1/2}I\approx 1.5$ for 
a cold disk [see eq. (\ref{Icold})], 
and is significantly smaller for hot disks 
[it is likely to be $\sim(m_0/M_p)^{2/3}$, see \S \ref{randmot}].

If we take planetesimals to be rocky bodies with radius $\sim 50$ km
and mass $\sim 10^{21}$ g,
then $m_0/(\Sigma_0 a^2)\sim 10^{-6}-10^{-7}$.
In the end we obtain that 
\begin{equation}
M_{crit}/M_{is}\sim (10^{-1}-10^{-2}).\label{resrat}
\end{equation}
This result means that 
accretion is slowed down long before the 
clearing of the feeding zone. We conclude
that the masses of
planetary embryos are $\sim 10$ to $100$ times smaller 
than predicted by arguments based on clearing the feeding zone 
[see eq. (\ref{isol})].

\subsection{Random motions of planetesimals}\label{randmot}
 
Equation (\ref{1pop}) fully describes the surface density evolution (or
its steady-state structure) only if it is supplied with information
about random motions of particles within the planetesimal disk,
which determine $P(h,\Delta h)$ and its moment $I$.

For a cold disk (zero velocity dispersion) $I$ is given by (\ref{Icold}).
In our case the planetesimal swarm
 is unlikely to be cold, because planetesimals will be
scattered by the planet, and also will scatter each other.
An important point to note here is that these two types of scattering
probably operate in quite different regimes. Indeed, the
 natural parameter
determining the heating regime is the ratio of the velocity dispersion to
the shear across the Hill radius. For the scattering by the 
planet this parameter
is
$S\sim v/\Omega R_H$, with $R_H=a(M_p/M_c)^{1/3}$
 while for the interaction
with other planetesimals it is $s\sim v/\Omega r_H$,
with $r_H=a(2 m_0/M_c)^{1/3}$, 
so that $s\sim S(M_p/m_0)^{1/3}$.
Thus, scattering by the planetary embryo could be in a shear-dominated 
regime, while
mutual scattering of planetesimals is practically always
in a dispersion-dominated one.

During a passage within a Hill radius from a massive body, a
planetesimal initially 
on a circular orbit gets a significant kick, so that its velocity dispersion
increases by $\sim \Omega R_H$. Thus, the part of the disk within
a Hill radius from the planetary embryo
 will be heated to $v\sim\Omega R_H$ corresponding
 to $S\sim 1$ in time
$\sim \Omega^{-1}(a/R_H)=\Omega^{-1}\mu_p^{-1/3}$.
Thus the stirring rate by planetary scattering is
\begin{equation}
\frac{ds^2}{dt}\bigg|_{pl}\sim\Omega\left(\frac{M_p^3}{m_0^2 M_c}
\right)^{1/3},~~~~~\mbox{for}~~~~~S\la 1,
\label{plstir}
\end{equation}
within a radial distance $R_H$ from the embryo.
The same kind of estimate could be derived for the excitation of the vertical
random motions.

This heating rate quickly leads to $s\gg 1$ 
 and, thus, the self-heating of the planetesimal population should be 
calculated
 in a dispersion-dominated regime. In particular the integral $I$ in
 equation
 (\ref{1pop}) describing
 the scattering of planetesimals by their mutual interactions should be 
calculated
 in this approximation.
 
Stewart \& Ida (2000) considered velocity stirring in the 
dispersion-dominated regime.
Their results demonstrate that the horizontal
stirring rate is
\begin{equation}
 \frac{ds^2}{dt}\bigg|_{self}=\Omega\frac{\Sigma_0}{m_0} 
r_H^2 \langle P_{VS}\rangle,
\label{selfheat}
\end{equation}
if the vertical velocity dispersion is of the same order as a horizontal 
one (as we normally expect),
and they provide a closed analytic form for the 
stirring coefficient $\langle P_{VS}\rangle$ 
as a function of planetesimal velocity dispersion.
The coefficient $\langle P_{VS}\rangle$ is defined 
to represent the change
in the square of the relative eccentricity [see H\'enon \& Petit (1986)],
averaged over the vertical and horizontal orbital phases,
and the 
velocity distribution of planetesimals (Ida, 1990; Stewart \& Ida, 2000):
\begin{eqnarray}
 \langle P_{VS}\rangle=\int 
\Delta e_u^2(e_u,i_u,h,\tau,\omega)f(e_u,i_u)de_u^2 di_u^2
 \frac{3}{2}|h|dh\frac{d\tau d\omega}
{4\pi^2},
\end{eqnarray}
where $e_u, i_u, \tau, \omega$ are the relative eccentricity, 
inclination, horizontal
and vertical orbital phases, $h$ is the separation of semimajor axes of the
 interacting particles and $f(e_u,i_u)$ is the corresponding distribution 
function.
 We assume
that $e_u$, $i_u$, and $h$ are properly normalized by the Hill radius for
particles participating in the collision [in this case our definition of 
$\langle P_{VS}\rangle$ differs from Stewart \& Ida (2000) by a factor of
$(m_1+m_2)^{4/3}/(3M_c)^{4/3}$, where $m_1$ and $m_2$ are the masses of 
interacting planetesimals]. 

Vertical stirring is described by a formula 
similar to
(\ref{selfheat}) but with a different stirring coefficient
\begin{eqnarray}
\langle Q_{VS}\rangle=\int \Delta i_u^2(e_u,i_u,h,\tau,\omega)
f(e_u,i_u)de_u^2 di_u^2
 \frac{3}{2}|h|d h
\frac{d\tau d\omega}{4\pi^2}.\label{oldps}
\end{eqnarray}

Now we can say more about the relation of the integral $I$ to the 
kinetic properties
of the planetesimal population.
One can easily see that the
averaging over all possible $\Delta h$ for
a given initial $h$ used in definition (\ref{Igen})
 is equivalent to averaging
over the vertical and horizontal orbital phases at a given
relative eccentricity and
inclination and then over the distribution of relative eccentricities and
inclinations, that is
\begin{equation}
I=\frac{1}{2}\int\limits_{-\infty}^{\infty}|h|dh\int
f(e_u,i_u)de_u^2 di_u^2\frac{d\omega d\tau}{(2\pi)^2}
[2h\Delta h +(\Delta h)^2].\label{inter1}
\end{equation}
From the conservation of Jacobi constant one has
$2h\Delta h +(\Delta h)^2=(4/3)[\Delta (e_u^2)+
\Delta ( i_u^2)]$.
Then, substituting into (\ref{inter1}) we get that
\begin{equation}
I=\frac{4}{9}\left(\langle P_{VS}\rangle +\langle Q_{VS}\rangle
\right).
\label{I}
\end{equation}
This is an important result, because it relates the evolution of the 
surface density of a spatially inhomogeneous planetesimal population
to the viscous stirring in a homogeneous disk.

Using these expressions we can determine the relative role of self-heating
of the disk and planetary heating
in the vicinity of the massive body. Of course, the former dominates beyond
several $R_H$ from planet, but within $\sim R_H$ of the embryo one can 
get from equations (\ref{crit}), (\ref{plstir}), (\ref{selfheat}), 
and (\ref{I}) the simple result that
\begin{eqnarray}
\frac{(ds^2/dt)|_{pl}}{(ds^2/dt)|_{self}}\sim
\lambda.\label{heatrat}
\end{eqnarray}
This means that for $\lambda \ga 1$, when a gap starts to form,
embryo perturbations begin to dominate planetesimal heating within
$\sim R_H$ of the planet, i.e. when the planet dominates the surface 
density evolution it also dominates the heating.

The evolution of kinetic properties of planetesimals could be neglected
if there is an effective velocity damping due to inelastic collisions
between bodies, as in the case of planetary rings (Petit \& H\'enon, 1987a),
or if there is a strong gas drag. In the planetesimal case, however, 
gravitational
stirring dominates over damping (Kenyon \& Luu 1998) 
and then the fact that a gap in the 
distribution of guiding centers is opened does 
not automatically
mean that planetesimals cannot reach the planet, because in the course of 
scattering
their velocity dispersion grows as well [see equation (\ref{heatrat})]. 
However, the accretion rate will 
drop anyway at least because of the less pronounced focussing 
(Safronov, 1972; Dones \& Tremaine, 1996). Also, inhomogeneous 
random velocity evolution or gas drag could remove the
residual planetesimals from the forming gap, thus bringing 
their surface density around the embryo to zero and shutting down 
accretion completely (see TI for N-body simulations in the presence 
of the gas drag).
For this reason we believe that our results with no velocity evolution 
are applicable to the problem of planet accumulation in many cases.

All the stirring coefficients are functions of the vertical and
horizontal velocity dispersions of planetesimal population.
Stewart \& Ida (2000) have in particular shown that $\langle P_{VS}\rangle, 
\langle Q_{VS}\rangle\propto s^{-2}\ln s$
 for $s\gg 1$ [we used this result 
 to derive equation (\ref{Ihot})].
It means that
to close properly 
the problem we need to couple equations for the velocity evolution
to equation (\ref{1pop}).
In doing so, one should
bear in mind that random motions are highly nonuniform in space,
since stirring by the massive body
 is strongly localized within several Hill radii
from it. Also, it is not clear that the velocity 
distribution  of planetesimals
can be adequately described by the Schwarzschild distribution, as is
usually assumed for simplicity.
For this reason we will not pursue this subject here and postpone its more
detailed exploration to a future work.

It is important to note however, that $I\propto s^{-2}\ln s$ and 
$\lambda\propto s^2/\ln s$
for $s\gg 1$, as
follows from the equations (\ref{Ihot}) and (\ref{hotcrit}). 
Thus, as $\lambda$ grows and the embryo heats up 
planetesimal population around it, the planetesimal ``viscosity'' 
decreases (increasing $\lambda$ even more through this velocity 
coupling), which facilitates gap
opening. This only strengthens our conclusion that a 
gap must form 
when the condition $\lambda\ga 1$ is fulfilled, even without knowing 
the details of the random velocity evolution of planetesimals.

\subsection{Effects of planetary migration.}\label{migration}

In \S \ref{res} we studied gap opening around a massive body, assuming
that the background distribution of surface density of planetesimals 
is symmetric with respect to the position of the embryo. In this case
we assumed that the 
embryo is fixed in radius and there is no migration at all.

It is more than likely that in real protoplanetary disks there
are significant surface density gradients, which could drive 
embryo migration. 
This could in principle introduce significant changes
into our picture. Indeed, if the embryo is able to migrate quickly it may
move out of  the gap it starts to form and, thus, gap formation
 would be suppressed.

Planetary migration will naturally occur in the course of 
embryo accumulation,
since in the process
of scattering planetesimals, the massive body
exchanges its angular momentum with them,
which leads to its migration. 
Indeed, planetesimals passing within $R_H$ from the embryo get 
displaced by a distance $\sim R_H$, thus an embryo itself is displaced
by $\sim R_H(m_0/M_p)$. During the time interval $\Delta t$ approximately
$(\Sigma_0/m_0) R_H^2 \Omega\Delta t$ planetesimals 
pass within embryo's Hill sphere. The 
surface number densities on both sides will likely be different by 
$\sim (\Sigma_0/m_0)(R_H/a)$, although it is likely that as the embryo
moves in some direction it plows planetesimals in front of it, leaving behind
a depression, which would tend to oppose the migration 
[see Ward \& Hourigan (1989) for a similar effect in gaseous disks]. 
Thus, our previous assumption about the surface density difference is likely
to be an upper limit and the actual migration will be weaker. 
One can easily calculate that the rate 
of this ``maximum'' migration is
\begin{equation}
\frac{1}{R_H}\frac{da}{dt}=
\frac{dh}{dt}\sim\Omega\frac{\Sigma_0 a^2}{M_c}.\label{selfmigr} 
\end{equation}
The time it takes the 
embryo to migrate through a zone with a width equal to  its own Hill
radius is then $\Omega^{-1}M_c/(\Sigma_0 a^2)$ and  should be longer
than $t_{open}$ given by equation (\ref{topen}) if a gap is to 
be maintained. Thus, the 
necessary condition here is
\begin{equation}
\left(\frac{M_p}{M_c}\right)^{1/3}>\frac{\Sigma_0 a^2}{M_c}.\label{cond1}
\end{equation}
If, say, $\Sigma_0 a^2/M_c=10^{-5}$, (and $\lambda\ga 1$) 
then all bodies with masses $\ga 10^{18}$ g
will open a gap faster than they migrate through it.

Another type of migration could arise if
the whole system is immersed in a massive gaseous disk, 
as should be the case in
the early stages of protoplanetary evolution.
In this case
Goldreich \& Tremaine (1980) demonstrated that it takes time 
$\sim \Omega^{-1}(h^2\Delta a/a^3)(M_c^2/M_p \Sigma_g a^2)$ for a planet to 
migrate a distance $\Delta a$ in the radial direction, where 
$h$ is the disk thickness, determined by its temperature, and
$\Sigma_g$ is the surface density of gaseous disk. In our case, the relevant
lengthscale is again $\Delta a \sim R_H$, thus migration timescale is
\begin{equation}
t_{mig}\sim\Omega^{-1}\mu_p^{-2/3}\frac{M_c}{\Sigma_g a^2}\frac{h^2}{a^2}.
\end{equation}
Comparing these two timescales we obtain:
\begin{equation}
\frac{t_{open}}{t_{mig}}\sim\mu_p^{1/3} \frac{\Sigma_g a^2}{M_c}
\frac{a^2}{h^2}\sim 10^{-3}
\left(\frac{M_p}{10^{24}~\mbox{g}}\right)^{1/3}
\frac{\Sigma_g a^2}{10^{-3}~M_\odot}
\left(\frac{M_c}{M_\odot}\right)^{-4/3}
\left(\frac{a/h}{30}\right)^{2}.
\end{equation}

Thus, migration due to the interaction with a gaseous disk
is unlikely to have an important effect on gap opening.

\section{Summary}\label{summary}

We studied the possibility that planetary formation due to the 
accretion of planetesimals could be significantly slowed or even stopped,
due to gap formation around a forming planetary embryo, caused by the strong
gravitational perturbations of planetesimals in its vicinity. 
We find a critical parameter $\lambda$
 which describes the importance of gap formation [eq. (\ref{crit})]. 
Predictions based on this parameter were compared with numerical
N-body simulations of this process (IM93, TI97),
and they are in good agreement.

Only the case of a single mass distribution of the particles
in a disk was studied here. But it is plausible 
that our basic results hold true even if a distribution
of masses exists. Only the characteristic planetesimal mass entering
this parameter should be chosen carefully, and this question merits further 
investigation [see Kokubo \& Ida (1996), (1998) for some numerical results]. 

Our findings were confirmed by solving the 
evolution equation (\ref{1pop}) neglecting the velocity dispersion
evolution of
planetesimals in the disk. The kinematic properties  of the planetesimal 
population are important in this sort of study, and the
 surface density evolution and velocity
evolution of planetesimals are closely related. 
Nevertheless, we hope to have grasped the
main qualitative features of the evolution of the 
distribution of guiding centers 
even without 
keeping track of the velocity evolution. The instantaneous density 
of planetesimals depends on their kinematic properties and should 
experience at 
least a decrease by a factor of several in the vicinity of the embryo, 
leading to slowing down the accretion 
(and gas drag and inhomogeneous velocity evolution could clear 
out the gap completely).
We also stress that our results for a cold disk provide an upper 
limit to the embryo mass required to open a gap.

If the disk is not uniform, 
migration of the embryo itself also likely to occur in the course of
its interaction with planetesimals or with the more massive gaseous disk from
which the whole embryo-planetesimal system originally condensed. 
We have estimated how this could 
affect our results and show that gap formation is likely to 
occur even when migration is present.

Finally, the evolution equation itself, coupled with the 
equations of planetesimal velocity evolution, provide us with 
a powerful tool to study the formation and evolution of planetary 
embryos. Since our 
results seem to be  in good agreement with N-body simulations, we may
use this machinery for other problems of a similar nature. 
An obvious example is the coupled evolution of several planetary embryos.

\section{Acknowledgements.}

I am grateful to my advisor, Scott Tremaine, 
for his encouragement and guidance,
and to J. Goodman and E. Chiang for fruitful discussions.
I would also like to acknowledge the financial support provided by 
NASA grants NAG5-7310 and NAG5-10456.


\appendix

\section{Derivation of equation (\ref{geneq}).}\label{ap1}
 
Following Petit \& H\'enon (1987b), we first calculate
the current of particles with mass $m_1$ (initially located at the 
guiding-center radius $r_1$) through the circle of
radius $r$,
due to the interaction with a single particle of mass $m_2$, located at 
guiding-center radius $r_2$.
The characteristic dimensionless relative distance for the two 
particles involved in this interaction
\begin{equation}
h=\frac{|r_1-r_2|}{r(\mu_1+\mu_2)^{1/3}}.\label{h}
\end{equation}
 It was demonstrated by H\'enon \& Petit (1986) that in coordinates
normalized in this way, the equations of relative motion of 
particles do not depend
upon their masses. For the particle $m_1$ initially located to the left of 
the boundary at $r$, to cross it to the right
one needs 
\begin{equation}
\Delta h>\Delta h_{min}=\frac{|r-r_1|}{r}\frac{m_1+m_2}
{m_2(\mu_1+\mu_2)^{1/3}},
\end{equation}
where the factor $(m_1+m_2)/m_2$ arises because $h$ describes relative
motion of particles.

Then the  left-to-right part of this flow of particles in the mass interval
$(m_1,m_1+dm_1)$ during the time
$dt$ is obviously given by
\begin{equation}
\int\limits^r_{-\infty}dr_1~ 2|A||r_1-r_2|\int\limits^\infty_{\Delta h_{min}}
d(\Delta h)\times 2\pi r_1 
P(h,\Delta h)N(m_1,r_1,t)dt ~dm_1. \label{rawflux}
\end{equation}
 Here $|A||r_1-r_2|$ is the local shear velocity between the interacting  
particles.

Summing over all possible positions and masses of particles $m_2$ we obtain
the total left-to-right flow of particles in the range $(m_1,m_1+dm_1)$:
\begin{eqnarray}
\langle\Delta J_+\rangle=dm_1\int\limits_0^{\infty}dm_2
\int\limits_{-\infty}^{\infty}dr_2\int\limits_{-\infty}^{r}dr_1
\int\limits_{\Delta h_{min}}^{\infty}d(\Delta h)\nonumber \\
\times
N(m_1,r_1,t)N(m_2,r_2,t)
P(h,\Delta h)4\pi r_1 |A||r_1-r_2|dt,\label{inteqn}
\end{eqnarray}
This formula coincides with equation (28) of PH.
We will further denote $f_i=f(m_i,...), i=1,2$ for {\it any}
function $f$ for brevity, and replace the factor $r_1$ under the integral
with $r$ (since $r_1$ weakly varies on the scale of the Hill radius).

Making the change of variables from $r_1, r_2$ to $R, h$ given by
[cf. equations (30) and (31) of PH]
\begin{eqnarray}
r_1=r+R,\\
r_2=r+R-(\mu_1+\mu_2)^{1/3} r h,
\end{eqnarray}
one can reduce equation (\ref{inteqn}) to
\begin{eqnarray}
\langle\Delta J_+\rangle=4\pi|A|r^3 dt ~
dm_1\int\limits_0^{\infty}dm_2
(\mu_1+\mu_2)^{2/3}
\int\limits_{-\infty}^{\infty}dh\int\limits_{-\infty}^{0}dR
\int\limits_{\Delta h_{min}}^{\infty}d(\Delta h)\nonumber \\
\times
N_1(r+R)N_2[r+R-(\mu_1+\mu_2)^{1/3} r h]P(h,\Delta h)|h|.\label{jplus}
\end{eqnarray}

We can change the order of integration over $d(\Delta h)$ and $dR$ to get
for $\langle\Delta J_+\rangle$
\begin{eqnarray}
\langle\Delta J_+\rangle= 4\pi |A|r^3 dt~ dm_1\int\limits_0^{\infty}dm_2
(\mu_1+\mu_2)^{2/3}
\int\limits_{-\infty}^{\infty}dh |h|
\int\limits_{0}^{\infty}d(\Delta h)P(h,\Delta h)
\int\limits_{D}^{0}dR\nonumber \\
\times
N_1(r+R)N_2[r+R-(\mu_1+\mu_2)^{1/3} r h],
\end{eqnarray}
with
\begin{equation}
D=-\frac{\mu_2 r}{(\mu_1+\mu_2)^{2/3}}\Delta h.
\end{equation}

Considering now the right-to-left flow of particles through $r$ 
one can get for this component of flux 
\begin{eqnarray}
\langle\Delta J_-\rangle=4\pi |A|r^3 dt~ dm_1\int\limits_0^{\infty}dm_2
(\mu_1+\mu_2)^{2/3}
\int\limits_{-\infty}^{\infty}dh |h|
\int\limits_{-\infty}^{0}d(\Delta h)P(h,\Delta h)
\int\limits_{0}^{D}dR\nonumber \\
\times
N_1(r+R)N_2[r+R-(\mu_1+\mu_2)^{1/3} r h].
\end{eqnarray}

Now, the total flux of particles through the boundary at $r$ is given by
$\langle\Delta J\rangle=\langle\Delta J_+\rangle-\langle\Delta J_-\rangle$.
Then the equation of evolution is obtained by setting
\begin{equation}
\frac{\partial}{\partial t}\left[2\pi r N_1(m,r,t)\right]dt  dm_1=
-\frac{\partial \langle\Delta J\rangle}{\partial r}. \label{differ}
\end{equation}

Here we do not differentiate $r^3$ in the right-hand side because it varies
only weakly on the scale of Hill radius.
Then the right-hand side of (\ref{differ}) 
contains $\partial (N_1 N_2)/\partial r$
which obviously equals $\partial (N_1 N_2)/\partial R$.
Taking this into account,
one can trivially perform an integration over $R$ in   the
r.h.s. of (\ref{differ}) to obtain finally equation (\ref{geneq}).
In deriving it we have also taken into  account that
\begin{equation}
\int\limits_{-\infty}^{\infty}d(\Delta h)P(h,\Delta h)=1.
\end{equation}

One should note that in deriving (\ref{jplus}) we assumed that
$\langle N_1 N_2\rangle_\phi=\langle N_1\rangle_\phi\langle N_2\rangle_\phi$,
where $\langle g \rangle_\phi$ means averaging quantity $g$ over the azimuthal 
angle. This might not be true in the planetary disks which are cold and
have a large viscosity (Spahn \& Srem\v{c}evi\'c 2000), or during the initial 
stages of the gap development in planetesimal disk. However, we believe that
it is unlikely to affect our results, since for hot disks and late times
this separability assumption should be adequate. Thus, it is possible 
that our numerical results presented in \S \ref{res} are somewhat different 
quantitatively at very early times
from what a more detailed theory would predict. But we
believe that our principal conclusions remain unchanged.

\section{Form of $h(h^\prime)$ used in PH.}\label{ap2}

Petit \& H\'enon (1987b) have solved numerically the Hill equations 
in the case when the initial random motion of interacting particles is small. 
In this case the outcome of the interaction between two particles is
deterministic, and they obtained the following form for the 
function $h(h^\prime)$, where
$h$ is the initial difference of semimajor axes of particles and $h^\prime$
is the final value of the same quantity, normalized by
$a[(m_1+m_2)/M_c]^{1/3}$, where $m_1$ and $m_2$ are the masses of interacting 
bodies:
\begin{eqnarray}
h^\prime(h)=\left\{
\begin{array}{ll}
h+3.34377(h^5+0.2 h^4-3.14 h^3)^{-1},  & \mbox{if}~~~h\geq 1.75622,\\
350 h^2-1204 h+1038.5,  & \mbox{if}~~~1.75622>h\geq 1.6777,\\
2895.903 h^5-20454.39 h^4+57671.78 h^3 &   \\
-81146.35 h^2+56984.57 h -15977.97,  & \mbox{if}~~~1.6777>h\geq 1.2219,\\
-1832.5 h^2+4361.35 h-2596.5,  & \mbox{if}~~~1.2219>h\geq 1.17,\\
-h-4.107085921(h^{-1}+4 h^4)\exp(-5.58505361 h^3),  
& \mbox{if}~~~1.17>h\geq 0.\\
\end{array}
\right.\label{phscat}
\end{eqnarray}

\end{document}